\documentclass[preprint,showpacs,preprintnumbers,amsmath,amssymb]{revtex4}

\usepackage{graphicx}
\usepackage{dcolumn}
\usepackage{bm}

\newcommand{\be}{\begin{equation}}
\newcommand{\ee}{\end{equation}}
\newcommand{\bea}{\begin{eqnarray}}
\newcommand{\eea}{\end{eqnarray}}

\begin{document}
\preprint{IMSc/2005/08/20}

\title{Gentile statistics and restricted partitions}

\author{C. S. Srivatsan}
\affiliation{Department of Physics,Indian Institute of Technology,Kanpur 
208016, India} 
\author{M. V. N. Murthy}
\affiliation{The Institute of Mathematical Sciences, Chennai 600 113, 
India}
\author{R. K. Bhaduri}
\affiliation{Department of Physics, McMaster University, Hamilton, Ont. 
L8S 4M1, Canada}
\date{today}

\begin{abstract}
In a recent paper (Tran et al., Ann.Phys.311(2004)204), some asymptotic
number theoretical results on the partitioning of an integer were
derived exploiting its connection to the quantum density of states of a
many-particle system. We generalise these results to obtain an
asymptotic formula for the {\it restricted or coloured} partitions
$p_k^s(n)$, which is the number of partitions of an integer $n$ into the
summand of $s^{th}$ powers of integers such that each power of a given
integer may occur utmost $k$ times. While the method is not rigorous, it
reproduces the well known asymptotic results for $s=1$ apart from
yielding more general results for arbitrary values of $s$.
\end{abstract}


\maketitle

\section{Introduction}

It is well known that in one-dimension, the density of states of a
system of ideal bosons confined in a harmonic trap at a given energy
$E$, in the $N\rightarrow \infty$ limit, is the same as the number of
ways of partitioning an integer $n=E$ into a sum of integers
\cite{grossman}. The result is the celebrated Hardy-Ramanujan
formula\cite{hardy}. An in-depth analysis of the fundamental additive
decomposition of positive integers by sums of other positive integers
with various restrictions is given in Ref.(\cite{book}).

In a recent paper Tran et al \cite{tran} exploited the connection
between quantum many particle density of states and integer partitions
in number theory to obtain more general results for the additive
decomposition or partitioning of an integer. It was shown that the
asymptotic density of states for ideal bosons in a system with a power
law energy spectrum, is analogous to $p^s(n)$, that is, the number of
ways of partitioning an integer $n$ as a sum of $s^{th}$ powers of
integers.  It was further shown that {\it distinct} partitions $d^s(n)$
correspond to the asymptotic density of states of a {\it pseudo
fermion}-like system in a power law energy spectrum. The name
pseudo fermion is coined to indicate that the system's density of
states was constructed by applying the exclusion principle only to the 
particle
distribution. The hole distribution was neglected while allowing any
number of particles in the ground state, i.e., a collapse of the
ground state, where the Fermi energy $E_F$ is set to zero. For the rest
of the paper we follow the same procedure as in \cite{tran} where only
particle distribution, relevant to partitions, is taken into account but
not the hole distribution. We note that the method is not rigorous but
yields correct asymptotic results in the well known cases.

In this paper we extend the analysis to obtain an asymptotic formula for
restricted or coloured partitions. We define $p_k^s(n)$ as the number of
partitions of an integer $n$ into a sum of $s^{th}$ powers of integers,
where each power of a given number occurs utmost k times. The method
used here is similar to the one used in the previous paper\cite{tran}.
The leading exponential approximation for the asymptotic partition of an
integer into integer summands, $s=1$, has been discussed recently by
Blencowe and Koshnick\cite{bk}.

We first construct the partition function for a statistical system in
which the maximal occupancy of each state is given by $k$. Obviously
$k=1$ corresponds to a fermionic system where as there are no
restrictions on $k$ for a bosonic system. Indeed the partition function
obtained by this minimal restriction is the same as that obtained by
Gentile\cite{gentile} and the corresponding statistics is know as {\it
Gentile Statistics}.  The partition function of Gentile statistics also
has the property that it nicely interpolates between the Fermi and Bose
Statistics. While the asymptotic formula, obtained using Gentile
statistics, reduces to the result for distinct partitions in the
fermionic, $k=1$, limit. The unrestricted or bosonic partitions are
obtained by taking the $k\rightarrow\infty$ limit.

\section{Many-particle density of states}

Before we proceed with the main theme of this paper, we recall briefly
results from the previous analysis for the sake of completeness.
Consider a system of $N$ particles, where $N$ is very large. In general,
the particles may obey any arbitrary statistics. The canonical
$N-$particle partition function is given by
\be
Z_N(\beta)~=~\int_0^{\infty}\rho_N(E) \exp(-\beta E) dE~,
\ee
where $\beta$ is the inverse temperature, ${E_i^{(N)}}$ are the
eigenenergies
of the $N-$particle system and $\rho_N(E)=\sum_i\delta(E-{E_i^{(N)}})$
is the $N-$particle density of states.  The density of states
$\rho_N(E)$ may therefore be obtained through the inverse Laplace
transform of the canonical partition function
\be \rho_N(E)
=\frac{1}{2\pi i} \int_{-i\infty}^{i\infty}\exp(\beta E) Z_N(\beta)
d\beta ~.
\label{rho1}
\ee
In general, it is not always possible to do this inversion analytically.
As in the case of single-particle density of states, the many particle
density of states may be decomposed into an average (smooth) part, and
an oscillating part~\cite{brack}. The smooth part $\overline{\rho}_N(E)$
may be obtained by evaluating Eq.(\ref{rho1}) using the saddle-point
method~\cite{kubo}. However, unlike the one-particle case, where the
oscillating part may be obtained using the periodic orbits in a ``trace
formula''~\cite{brack}, it remains a challenging task to find an
expression for the oscillating part $\delta\rho_N(E)$~\cite{sakhr}. In
what follows we shall use the saddle-point method to obtain the smooth
part of the density of states, $\overline{\rho}_N(E)$ and identify it
with the number of ways the energy $E$ is partitioned, on the average,
among $N$ particles.

Defining the entropy, $S(\beta)$ as
\be
S(\beta) = \beta E + \log Z_N .
\label{entropy}
\ee
and expanding the entropy around the stationary point $\beta_0$ and
retaining
only up to the quadratic term in Eq.(\ref{rho1}) yields
the standard result~\cite{kubo}
\be
\overline{\rho}_N(E) = \frac{exp[S(\beta_0)]}{\sqrt{2\pi S''(\beta_0)}},
\label{rho2}
\ee
where the prime denotes differentiation with respect to inverse
temperature and
\be
E=-\left( \frac{\partial \ln Z_N}{\partial \beta}\right)_{\beta_0}.
\label{saddle}
\ee

The above analysis holds for any $N$ particle system irrespective of the
nature of the single particle spectrum. Here after we omit the the
subscript $N$ since we are only interested in the asymptotic limit
$N\rightarrow \infty$.

\section{Restricted Partitions}

We consider a statistical system in which a given quantum state has a
maximal occupancy of $k$ in each single particle state. Thus when $k=1$
the system corresponds to a system of fermions whereas there are no such
restrictions on $k$ for a system of bosons except for symmetry
considerations. We now construct the partition function for such a
system which also incorporates the property of interpolation between
Bose and Fermi statistics as $k$ takes on different values.

In order to relate to the problem of integer partitions, we consider a
system with a single particle spectrum given by $\epsilon_m
=m^s,$ where the integer $m\geq 1$, and $s>0$. The energy is rendered
dimensionless by appropriate choice of units. For example, when
$s=1$ the spectrum can be mapped on to the spectrum of a one dimensional
oscillator, where the energy is measured in units of $\hbar\omega$.  For
$s=2$, it is equivalent to setting energy unit as $\hbar^2/2M$, where
$M$ is the particle mass in a one dimensional square well with unit
length.  These are the only two physically interesting cases. As in
\cite{tran}, we however keep $s$ arbitrary even though for $s >2$ there
are no quadratic Hamiltonian systems.

We define a {\it Restricted or Coloured Partition} of an integer $n$ as
its additive
decomposition into integer powers where no one integer is repeated more
than $k$ times. For example consider partitions of an integer, say $N=6$
as a summand of integers:
\begin{eqnarray}
6,~5+1,~4+2,~3+2+1, & &~( p_1^1(6)=4 ) \nonumber\\
3+3,~4+1+1,~2+2+1+1, & &~( p_2^1(6)=7 ) \nonumber\\
3+1+1+1,~2+2+2, & &~( p_3^1(6)=9 ) \nonumber\\
2+1+1+1+1, & &~( p_4^1(6)=10 ) \nonumber \\
1+1+1+1+1+1, & &~(p_6^1(6)=11)~.
\end{eqnarray}
We are interested in the relation between the asymptotic
expression for such restricted partitions and the density of states of
the
system which allows utmost $k$ occupancy of the single particle states.

The relevant physical system for obtaining partitions is thus one in
which at zero temperature all the particles are in the ground state
whose energy is set to zero. At any excitation energy, particles are
excited from the ground state and distributed in the single particle
states with maximal occupancy given by $k$ in each state. The number of
ways in which such a distribution can be achieved is the density of
states at that energy (as also the number of ways of partitioning the
energy).

The {\it partition function} for such a system in the limit of the
number of particles
$N\rightarrow \infty$ is given by
\bea
Z_{\infty}(\beta) &=&
\prod_{m=1}^{\infty}[1+\exp(-\beta m^s) + \cdots +\exp(-k\beta
m^s)]\nonumber\\
&=&~\prod_{m=1}^{\infty}~\sum_{n=0}^{k}\exp(-n\beta m^s),
\label{zprod}
\eea
where the single particle spectrum is given by a power law. We note that
the
partition function in Eq.(\ref{zprod}) is indeed the grand-canonical
partition function with the chemical potential $\mu=0$. In particular
$k=1,\infty$ correspond to the two cases considered in the earlier
paper\cite{tran}. Note that with $k$ arbitrary this is indeed
the partition function for the well-studied Gentile
statistics\cite{gentile}.

By setting $x=\exp(-\beta)$, the partition function
may be written as
\begin{equation}
Z_{\infty}(x)=\sum_{n=1}^{\infty} p_k^s(n) x^n
=\prod_{n=1}^{\infty}\frac{[1-x^{(k+1)n^s}]}{[1-x^{n^s}]}.
\label{sis}
\end{equation}
In the asymptotic limit of large number of particles, $p_k^s(n)$ is the
number of ways of partitioning $n$.

Using Eq.(\ref{entropy}),we obtain
\be
S=\beta E+\sum_{n=1}^{\infty}~\ln[1-\exp(-(k+1)\beta
n^s)]-\sum_{n=1}^{\infty}~\ln[1-\exp(-\beta n^s)].
\label{sinfty}
\ee

We now evaluate the sums approximately using the Euler-Maclaurin
series. We have
\begin{eqnarray}
&-&\sum_{n=1}^{\infty} \ln[1-\exp(-\beta n^s)]=-\sum_{n=0}^{\infty}
\ln[1-\exp(-\beta (n+1)^s)]\nonumber\\
&=&
-\int_{0}^{\infty} \ln[1-\exp(-\beta (x+1)^s)]dx
-\frac{1}{2}\ln[1-\exp(-\beta)]
+s\sum_{k=1}^{\infty} \frac{B_{2k}}{2k(2k-1)},
\label{em}
\end{eqnarray}
where $B_{2k}$ are the Bernoulli numbers.
The integral in the Euler-Maclaurin expansion may
be evaluated as follows:
\begin{eqnarray}
&-&\int_{0}^{\infty} \ln[1-\exp(-\beta (x+1)^s)]dx \nonumber\\
&=&-\int_{0}^{\infty} \ln[1-\exp(-\beta x^s)]dx
+\int_{0}^{1} \ln[1-\exp(-\beta x^s)]dx \nonumber \\
&=&
\frac{C(s)}{\beta^{1/s}}
+\ln[1-\exp(-\beta)] - s\int_{0}^{1} \frac{\beta x^s}{\exp(\beta x^s)
-1}dx,
\end{eqnarray}
where
\begin{equation}
C{(s)} = \Gamma(1+\frac{1}{s})
\zeta(1+\frac{1}{s})
\end{equation}
given in terms of the Riemann zeta function. In the {\it high
temperature limit}, that is $\beta \rightarrow 0$, we have
\begin{equation}
-\int_{0}^{\infty} \ln[1-\exp(-\beta (x+1)^s)]dx
\approx \frac{C(s)}{\beta^{1/s}}
+\ln[1-\exp(-\beta)] - s + O(\beta)
\end{equation}

The series involving Bernoulli numbers may be evaluated approximately as
follows: The Euler-Maclauring series applied to logarithm of Gamma
function gives,
\begin{equation}
\ln (\Gamma(n+1)) =n\ln(n) -n +\frac{1}{2}\ln(2\pi) +\sum_{k=1}^{\infty}
\frac{B_{2k}}{2k(2k-1) n^{2k-1}}
\end{equation}
We may consider this as the expansion of zero by putting $n=1$ and get
\begin{equation}
s\sum_{k=1}^{\infty} \frac{B_{2k}}{2k(2k-1)}
=s - \frac{s}{2}\ln(2\pi).
\end{equation}

Thus keeping terms up to $O(\beta)$, and using the above arguments the
last term in Eq.(\ref{sinfty}) may be written as
\begin{equation}
-\sum_{n=1}^{\infty}[1-\exp(-\beta n^s)]
\approx \frac{C(s)}{\beta^{1/s}}
+\frac{\ln[1-\exp(-\beta)]}{2} -\frac{s}{2}\ln(2\pi)+O(\beta)~
\label{euler}
\end{equation}
One may also get the above result directly using symbolic math program
MAPLE.
The infinite sum in the second term of Eq.(\ref{sinfty}) is also
evaluated similarly by scaling
$\beta$ appropriately. Note that the constant term does not get scaled.

Thus the entropy to the leading order in $\beta$ is given by
\be
S=\beta E + \frac{\alpha C{(s)}}{\beta^{1/s}}- {1\over
2}~\ln[1-\exp(-(k+1)\beta)] +{1\over 2}~\ln[1-\exp(-\beta)]+O(\beta)~,
\label{sinftyexp}
\ee
where
\be \alpha = 1-\frac{1}{(k+1)^{1\over s}}.
\label{alpha}
\ee
We note that the expansion in powers of $\beta$ is equivalent to taking
the high-temperature limit though this limit should be employed
cautiously when $\beta$ is weighted by $k$ which can in principle take
large values.  The last term of $ O(\beta)$ in the RHS of
Eq.(\ref{sinftyexp}) causes a shift in the energy E by a constant (for
example 1/24 in the s=1 case) in $S$. In the asymptotic formula for the
level density for large $E$, obtained by the saddle point method, this
may be ignored. Note, however, that such a shift was included by
Rosenzweig\cite{rosen} to improve the Bethe formula\cite{bethe} for the
nuclear level density. We further note that the $k\rightarrow \infty$,
or the bosonic limit, is special since this limit involves precisely the
sum given in Eq.(\ref{euler}) and not the difference of two series as in
Eq.(\ref{sinfty}). As a result the $\beta$ independent term, namely
$-(s/2)\ln(2\pi)$, appears only in this limit and not for any finite
$k$. Indeed, as we shall see below this term is crucial in getting the
prefactor correctly in the Hardy-Ramanujan Formula.

The saddle point on the real axis\footnote{Note that we evaluate the
saddle point only along the real axis. In general there may be other
complex saddle points that are ignored. We assume, without proof, that
this is the only one that gives
the correct asymptotic behaviour as it is known to yield the
rigorous
answers for ordinary partitions.} is evaluated by taking the derivative
of the entropy,
\be
S^{\prime}(\beta)=E-{1\over s}~ {\alpha C{(s)}\over
{{\beta^{(1+1/s)}}}}~,
\ee
where only the leading terms in $\beta$ are retained.
Equating the derivative to zero, the saddle-point in $\beta$ is given by
\be
\beta_0 = \left( \frac{\alpha C{(s)}}{s E}\right)^{s \over (1+s)}~
=~\kappa_s~ E^{-{s\over {1+s}}},
\label{betz}
\ee
where
\be
\kappa_s=\left({\alpha C{(s)}\over s}\right)^{{s\over {1+s}}}~.
\ee

Substituting this in the saddle point expression for the density of
states in Eq.(\ref{rho2}), the asymptotic density of states is given by
\be \overline{\rho}_{k}^{(s)}(E)~\approx~ \kappa_s\sqrt{s}
\frac{\exp\left[\kappa_s
(s+1)E^{{1\over
{1+s}}}\right]}{\sqrt{2\pi(s+1)E^{(3s+1)/(s+1)}[1-\exp(-(k+1)\kappa_s
E^{-s/(s+1)})]}}~,
\label{rhofty}
\ee
where the subscript $k$ in $\overline{\rho}$ indicates the main property
of the statistical system under consideration. We have kept the
exponential term under the square-root sign in the denominator to
indicate the interpolation property of the density of states between the
Bose ($k\rightarrow \infty$) and the Fermi ($k=1$) limits even though
keeping the exponential may not be consistent with the order of the
expansion. The correct expression is given below.

For finite $k$ there always exists an energy $E$ which is large enough
such that the following approximation is useful:
\be \overline{\rho}_{k}^{(s)}(E)~\approx~ \sqrt{s\kappa_s}
\frac{\exp\left[\kappa_s
(s+1)E^{{1\over
{1+s}}}\right]}{\sqrt{2\pi(s+1)(k+1)E^{(2s+1)/(s+1)}}}~,
\label{rhofty1}
\ee
Note that the above expression is identical to the Eq.(23) for $d_s(n)$
in \cite{tran} corresponding to the special case of $k=1$. The
degeneracy or $k$ dependence in the above equation is also hidden in the
parameter $\kappa_s$.
An asymptotic formula derived in Ref.(\cite{bk}) based on Gentile
statistics for the special case of $s=1$ is the same as the one given in
Eq.(\ref{rhofty1}). However, to the best of our knowledge, no general
formula for arbitrary $s$ and $k$ exists.

In particular for $s=1, k=1$ the above equation reduces to the well
known formula\cite{book,abramowitz} for  distinct partitions of an
integer into a set of integers
\be
\overline{\rho}(E)
\approx~d(n=E)~=\frac{\exp\left[\pi \sqrt{E\over 3}
\right]}{4\times 3^{1/4}E^{3/4}}~.
\ee

The bosonic limit may be
obtained from Eq.(\ref{rhofty}) by taking the limit $k\rightarrow
\infty$. However we also note that this limit should be taken in such a
way that the $\beta$ independent factor in Eq.(\ref{euler}) is also
included. With this proviso we obtain
\be
\overline{\rho}_{k}^{(s)}(E)~\approx~ \kappa_s\sqrt{s}
\frac{\exp\left[\kappa_s (s+1)E^{{1\over
{1+s}}}\right]}{(2\pi)^{(s+1)/2}\sqrt{(s+1)E^{(3s+1)/(s+1)}}}~,
\label{rhofty2}
\ee
which is the celebrated Hardy- Ramanujan asymptotic formula for integer
partitions. In particular for $s=1$, we have the well known result(see
for example Ref.(\cite{abramowitz}).

\be
\overline{\rho}(E)
~\approx~p(n=E)~=\frac{\exp\left[\pi \sqrt{2E\over 3}
\right]}{4\sqrt{3}E}~.
\ee

\begin{figure}[htp]
\vskip 13 truecm
\includegraphics{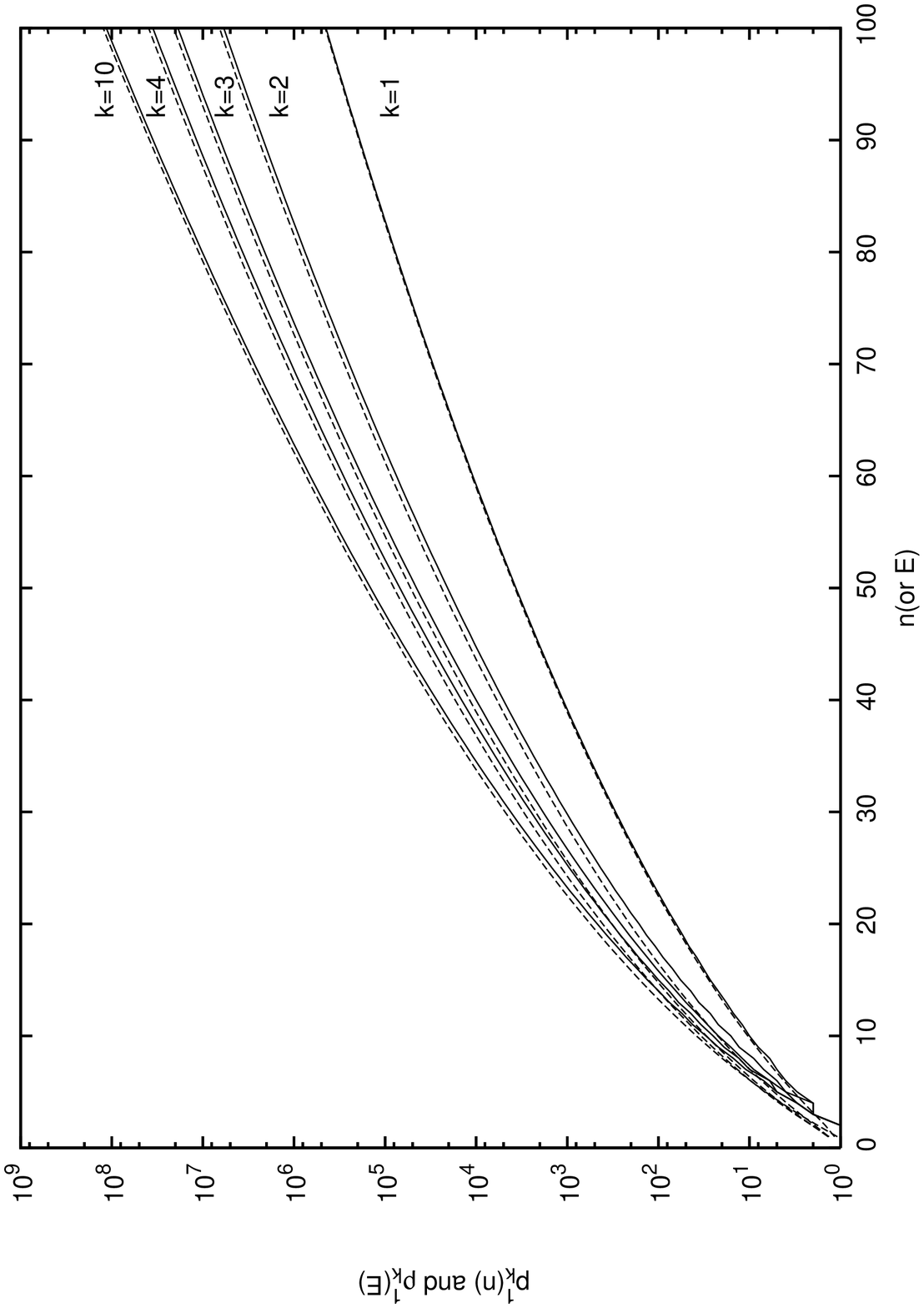}
\caption{Comparison of the exact $p_k^1(n)$ (solid line) and the 
asymptotic $\overline{\rho}^1_{k}(n)$ (dashed line) for $s=1$} 
\label{fig1}
\end{figure}

In Fig.~(\ref{fig1}) we show a comparison between the exact $p_k(n)$
(continuous line), and the asymptotic density of states,
$\overline{\rho}_{k}^{(1)}(n)$ (dashed line) obtained from
Eq.(\ref{rhofty1}) for various values of $k$ when the power $s=1$. For
smaller $k$ the asymptotics is reached earlier than for larger $k$ as
can be seen easily from the figures. In fact for $k=1$ the formula is
almost exact even for $n$ small. We note that the numerical estimates
for the density based
on Eq.(\ref{rhofty}) are usually higher than the exact values and
asymptotically reach the exact value later than the density calculated
based on Eq.(\ref{rhofty1}) which is shown in the figure. The saddle
point evaluated by keeping only the two leading terms in $\beta$ ensures
that the Eq.(\ref{rhofty1}) is more accurate for small values of $k$
though asymptotically both equations yield the same result.

\begin{figure}[htbp]
\vskip 13 truecm
\includegraphics{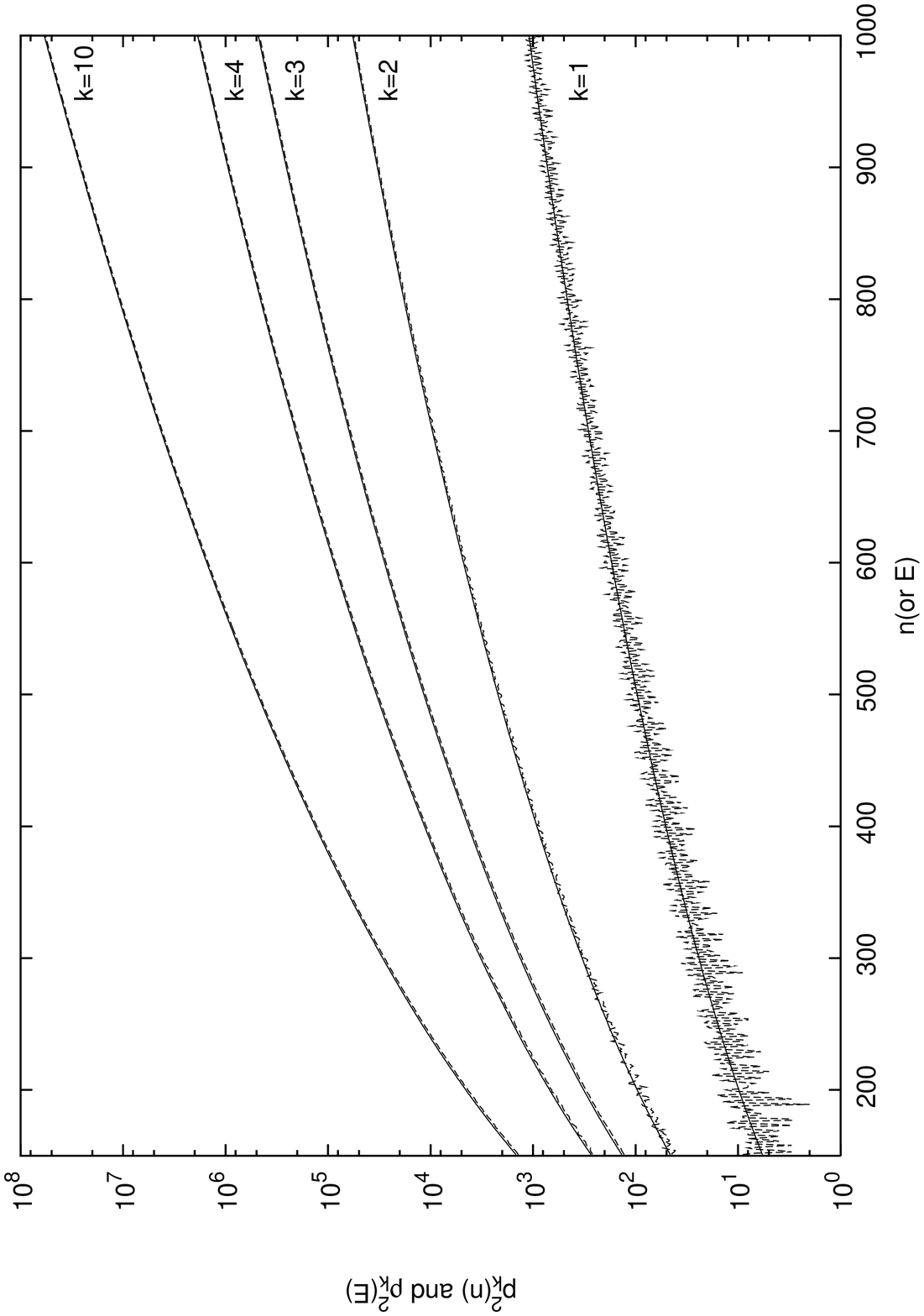}
\caption{Comparison of the exact $p_2^k(n)$ (solid line) and the 
asymptotic $\overline{\rho}^2_{k}(n)$ (dashed line) for $s=2$.}
\label{fig2} 
\end{figure}

\begin{figure}[htbp]
\vskip 13 truecm
\includegraphics{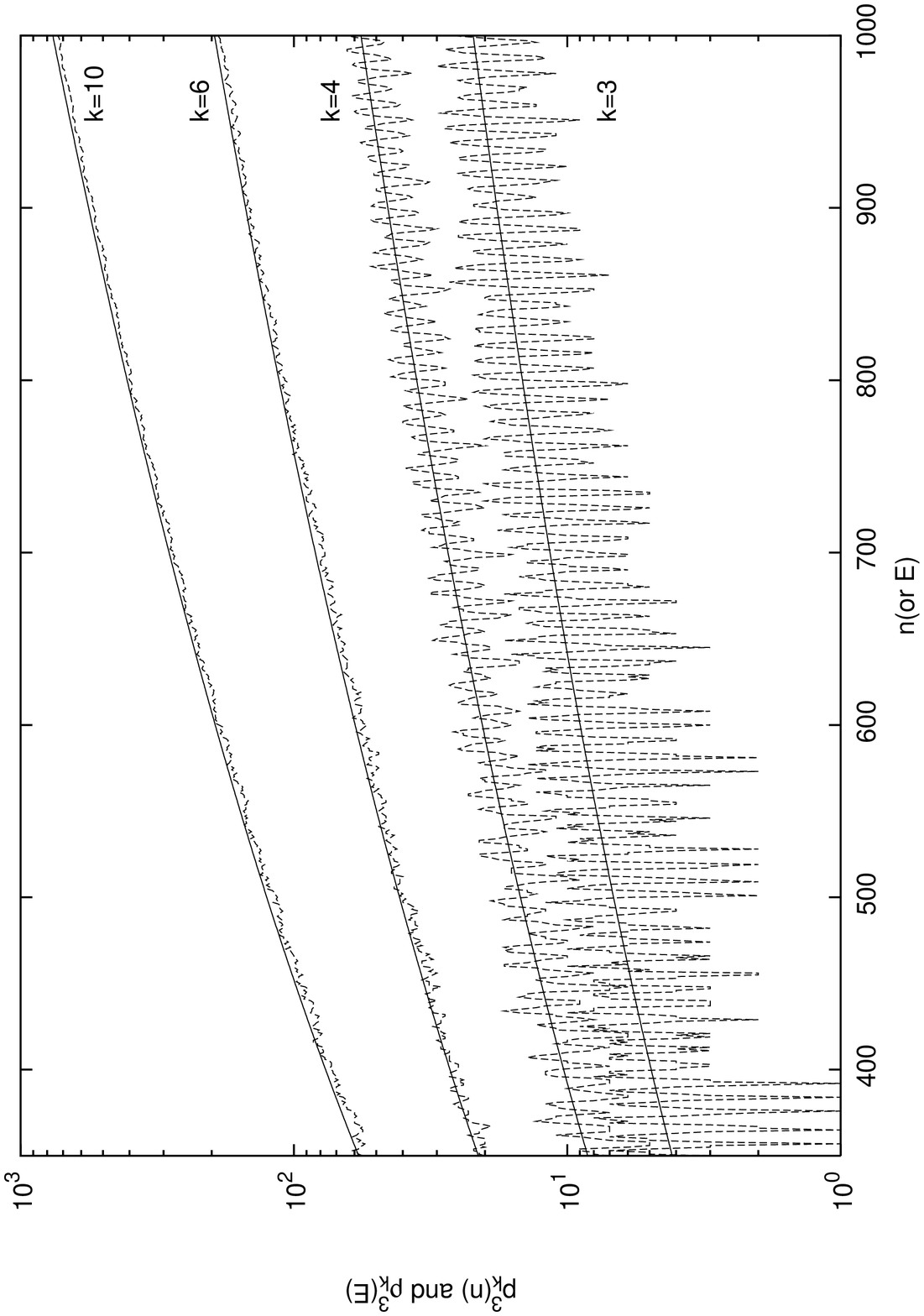}
\caption{Comparison of the exact $p_3^k(n)$ (solid line) and the 
asymptotic 
$\overline{\rho}^2_{k}(n)$ (dashed line) for $s=3$.}
\label{fig3} 
\end{figure}

Similarly, in Fig.~(\ref{fig2}), the computed $p_k^2(n)$ is compared
with $\overline{\rho}^{2}_{k}(n)$. We note that in this case the
fluctuation about the mean is larger for small values of $k$. In general
the fluctuations decrease with increasing $k$ for a given $s$ but
increase with increasing $s$ for a given $k$ as shown in Fig.\ref{fig3}
where we have plotted the asymptotic density of states for $s=3$ and
various values of $k$.

\section{Summary}

To summarise, this paper extends the results of Ref.  \cite{tran} to the
case of restricted or coloured partitions. We identify the relevant
generating function for coloured partitions as the partition function of
gentile statistics. The resulting asymptotic or smooth density states
interpolates between the Bose (arbitrary partitions) and Fermi (distinct
partitions) systems as a function of the maximal occupancy of the
quantum state.

We thank the referee for perceptive comments. This work was started when
one of us (CSS) was visiting IMSc as a summer research student in 2004. 
CSS and RKB would like to thank the hospitality of the Institute of
Mathematical Sciences where this work was done. RKB would like to
acknowledge financial support from NSERC (Canada).

\newpage
\newpage
\newpage

\end{document}